\documentclass[prd,showpacs,preprintnumbers,twocolumn,amsmath,nofootinbib,amssymb,floatfix]{revtex4}
\usepackage{graphicx,color,dcolumn,booktabs,bm}
\usepackage{epstopdf}
\usepackage{longtable,lscape}
\usepackage{txfonts}
\usepackage{overpic}
\usepackage{float}
\usepackage{amssymb}
\usepackage{amsmath}
\usepackage{epstopdf}
\usepackage{indentfirst}
\usepackage{feynmf}  
\usepackage{slashed}  
\usepackage{cases}
\usepackage{ulem}
\usepackage{color}
\usepackage{xcolor}
\usepackage{float}
\usepackage{multirow}
\usepackage{tikz}
\usepackage{epstopdf}
\usepackage{epsfig,dsfont,amssymb,amsmath,amsfonts,amsbsy,mathrsfs}
\usepackage{multirow}
\usepackage{graphicx}  
\usepackage{float}  
\usepackage{subfigure}  
\usepackage{array}
\usepackage{microtype}

\usepackage[colorlinks, citecolor=blue,anchorcolor=red,menucolor=red, linkcolor=red,filecolor=red,runcolor=red,urlcolor=blue,frenchlinks=true]{hyperref}
\makeatletter
\@addtoreset{equation}{section}
\makeatother

\newcommand{\nc}{\newcommand}
\nc{\tj}[1]{\textcolor{red}{Tianjin: #1}}

\usepackage{soul}
\allowdisplaybreaks
\begin{document}
\title{Newly observed $\kappa$(2600) and the highly excited states of the $K_0^{*}$ family}
\author{Cheng-Qun Pang$^{1,2}$}\email{xuehua45@163.com}
 \author{Hao Chen$^{3}$}\email{chenhao\_qhnu@outlook.com}
 \author{Yun-Hai Zhang$^{4}$}\email{wlxzyh@163.com}

\affiliation{
$^1$School of Physics and Optoelectronic Engineering, Ludong University, Yantai 264000, China
\\$^2$Lanzhou Center for Theoretical Physics, Key Laboratory of Theoretical Physics of Gansu Province, Lanzhou University, Lanzhou, Gansu 730000, China,
\\
$^3$College of Physics and Electronic Information Engineering, Qinghai Normal University, Xining 810000, China
 \\$^4$ College of Physics and Electronic Engineering, Heze University, Heze 274015, China
}

\date{\today}

\begin{abstract}
We conducted a study using the modified Godfrey-Isgur quark model and quark pair creation model to investigate the spectrum and two-body strong decays of the newly discovered $\kappa$(2600) resonance by the LHCb collaboration. Our analysis revealed that this {resonance} can be assigned as the fourth radial excitation within the $0^{+}$ light strange meson family. Furthermore, we predicted additional radial excitations in this meson family and systematically discussed their spectra, decay behaviors, and branching ratios. These results provide critical guidance for future to identify their signatures in experiments.
\end{abstract}
\maketitle

\section{introduction}\label{sec1}

Recently, the LHCb Collaboration reported the discovery of a new resonance $\kappa$(2600) (or named $K_0^*(2600)$ 
in the further states of  Particle Data Group (PDG)  \cite{ParticleDataGroup:2024cfk}) in the $B^+ \to (K_S^0 K \pi) K^+$ decay channel, with a statistical significance of 26.6$\sigma$, by analyzing the invariant mass spectrum of $K\pi$. The new resonance has {a mass of} $2662\pm59\pm201$ MeV {(the natural unit system is adopted in this work, where the speed of light $c=1$, and all mass units in the following text are expressed in MeV)} and {a width of} $480\pm47\pm72$ MeV, respectively \cite{LHCb:2023evz}.
This discovery intrigued us to investigate whether the $\kappa$(2600) could belong to the $J^{P}=0^{+}$ $K$ mesons family.

$P$-wave $K$ mesons have three $J$ numbers under the coupling of spin and orbital angular momentum, $^{2S+1}L_J$=$^1P_1$, $^3P_0$, and $^3P_2$. Li {\it{et~al}}. discussed a new structure  $X(2085)$ in $J^P=1^{+}$ $K$ meson family, and investigated the other members of this meson family \cite{Li:2024hrf}. In our previous work, we studied the $K_2^*$ meson family, especially the $K_2^*(1980)$ \cite{Li:2022khh}. For the low-lying excited states of $0^{+}$ light strange meson family, $K_0^*(1430)$ is the ground state, and $K_0^*(1950)$ can be its first excited state  \cite{Barnes:2002mu,Pang:2017dlw}. For the lowest scalar $\kappa(700)$, it is widely believed that it may be an exotic state \cite{Jaffe:2004ph,tHooft:2008rus,Close:2002zu,Parganlija:2012gv,Pelaez:2003dy,Eichmann:2015cra}. Wang {\it{et~al}}. identify that $K_0^*(2130)$ can be assigned as the second excited state of $0^{+}$ light strange meson family \cite{Li:2022ybj}. In Ref. \cite{Li:2022ybj},  {the predicted mass and width of the $K_0^*(4P)$ state are}  2404 MeV and 180 MeV, respectively, whereas its mass was 2424 MeV in our previous work \cite{Pang:2017dlw}. It is obvious that $\kappa$(2600) could not be $K_0^*(4P)$ state. The internal structure of this state, as well as the nature of the highly excited states of  the $0^{+}$ light strange meson family, becomes an interesting issue. 

Regge trajectories are an efficient method for investigating the spectrum of meson families \cite{Chew:1962eu,Anisovich:2000kxa}.
The {following} relation holds for the masses and radial quantum numbers of mesons within the same meson family
\begin{eqnarray}
M^2 = M_0^2 + (n - 1)\mu^2, \label{rt}
\end{eqnarray}
\begin{figure}[htbp]
\includegraphics[scale=0.5]{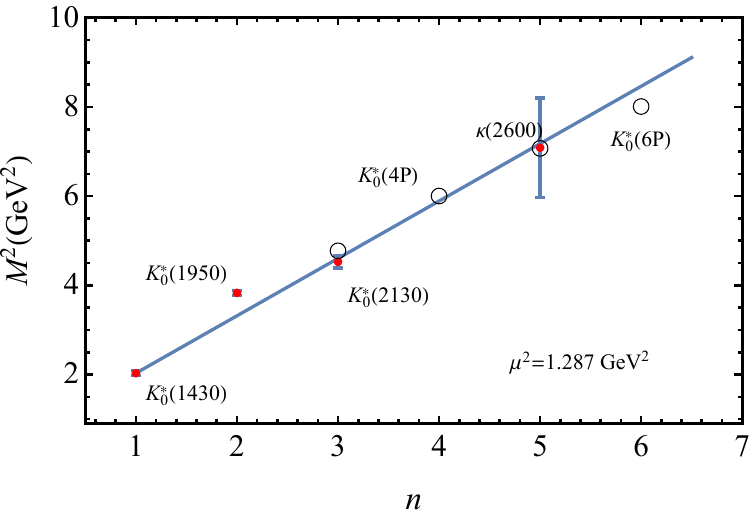}
\caption{The Regge trajectories {for the $K_0^*$ family. Here, open circles represent theoretical values (obtained with the modified Godfrey-Isgur model), and filled geometries indicate experimental  {measurements} \cite{ParticleDataGroup:2024cfk}. The radial quantum number, denoted by $n$, corresponds to the meson in question.}}
\label{regge}
\end{figure}
where $M_0$ denotes the mass of the ground state, $n$ represents the radial quantum number of the corresponding meson with the mass $M$, and $\mu^2$ is the trajectory slope. 
For the $K_0^*$ family, we adopt $\mu^2=1.287$ $\text{GeV}^2$ as depicted in Fig.~\ref{regge}.


By analyzing the Regge trajectory, the results demonstrate that $K_0^*(1430)$, $K_0^*(1950)$, and  $K_0^*(2130)$ are {the $1P$, $2P$, and $3P$ states}, respectively. The {newly} observed state, $\kappa(2600)$  can be a good candidate {for} the fourth  radially excited state of $K_0^*(1430)$. The $K_0^*(4P)$ is still misty. The mass of $K_0^*(4P)$ given by the modified Godfrey-Isgur (MGI) model is consistent with the value from the Regge trajectory and the mass of $K_0^*(6P)$ provided by the MGI model is slightly smaller than that given by the Regge trajectory.

{Both mass spectral analysis and two-body strong decay calculations provide effective approaches to probe the internal structure of mesons.} Via the simple mass {analysis, we find} that  $\kappa(2600)$  can be arranged as the fourth radially excited state of $K_0^*(1430)$. This preliminary conclusion needs to be substantiated with more quantitative analysis. We will study the mass spectra of the $0^{+}$ light strange meson family  using the  {MGI model}. 
In 1985, Godfrey and Isgur proposed the Godfrey-Isgur (GI) quark model for describing relativistic meson spectra with great success, exactly for low-lying mesons \cite{Godfrey:1985xj}, and for excited states, the screened potential must be taken into account for coupled-channel {effects} \cite{vanBeveren:2003kd,Liu:2009uz,2008contribution,Pang:2018gcn,Wang:2022xxi,Wang:2021abg, Wang:2024yvo,Feng:2022hwq}. 
We also study the {two-body} strong decays of the $0^{+}$ light strange meson family through {the} quark pair creation model (QPC {model}, also known as $^3P_0$ model) which was initially {formulated} by Micu \cite{Micu:1968mk} and is widely applied to the Okubo-Zweig-Iizuka (OZI)-allowed two-body strong decays of mesons in Refs. \cite{vanBeveren:1982qb, Titov:1995si, Ackleh:1996yt, Blundell:1996as, Bonnaz:2001aj, Zhou:2004mw, Lu:2006ry, Zhang:2006yj, Luo:2009wu, Sun:2009tg, Liu:2009fe, Sun:2010pg, Rijken:2010zza, Ye:2012gu, Wang:2012wa, He:2013ttg, Sun:2013qca,Pang:2018gcn, Wang:2022juf, Wang:2022xxi, Li:2022khh, Li:2022bre, Wang:2020due, Pang:2017dlw, Wang:2024yvo, Wang:2021abg, feng:2021igh, Feng:2022hwq}.

The paper is organized as follows: In Sec.~\ref{sec2}, we briefly review the Regge trajectory, the MGI model, and the $^3P_0$ model.
In Sec.~\ref{sec3}, the obtained masses of the $K_0^*$ family are presented and compared with available experimental data and other results from different {models}. Then, we systematically study the OZI-allowed two-body strong decay behaviors of the newly observed $\kappa(2600)$ and {predicted} states $K_0^*(4P)$ and $K_0^*(6P)$. Finally, a conclusion is given in Sec.~\ref{sec4}.

\section{Models employed in this work} \label{sec2}

In this work, we employ two models, the MGI model and the QPC model, which will be introduced in this section. {Taking} the  process  $K_0^*(2P)\to K^*\phi$ as an example, {the parameter dependence in the QPC model will be analyzed.}
 
\subsection{The modified GI model}

In 1985, Godfrey and Isgur proposed the  {GI} quark model for describing relativistic meson spectra with great success, {especially} for low-lying mesons \cite{Godfrey:1985xj}.
To describe the excited states well, the screened potential term {has been introduced} \cite{vanBeveren:2003kd,Liu:2009uz,2008contribution,Pang:2018gcn,Wang:2022xxi,Wang:2021abg, Wang:2024yvo,Feng:2022hwq}. Then, Song {\it{et~al}}. {proposed} the MGI model \cite{Song:2015nia} based on the GI model.
And it has {achieved} great success in calculating light {meson} spectroscopy \cite{Pang:2018gcn, Wang:2022xxi, Wang:2021abg, Wang:2024yvo, Feng:2022hwq}. In the MGI model, the Hamiltonian reads 
\begin{equation}\label{hh}
\widetilde{H}=\sum_i{({m_i^2+\mathbf{p}_i^2})}^{1/2}+\widetilde{V}^{\mathrm{eff}},
\end{equation}
where $m_i$ denotes the mass of the quark or antiquark. The masses adopted here are $m_s=0.377$ GeV and $m_{u(d)}=0.162$ GeV, which are smaller than those in the GI model ($m_s=0.419$ GeV, $m_{u(d)}=0.22$ GeV)  \cite{Godfrey:1985xj}, and $\widetilde{V}^{\mathrm{eff}}$ includes short-range $\gamma^{\mu}\otimes\gamma_{\mu}$ one-gluon-exchange interaction and $1\otimes1$ linear confinement interaction between $q$ and $\bar{q}$, which is given by
\begin{equation}\label{V}
\widetilde{V}^{\mathrm{eff}}=\widetilde{G}_{12}+\widetilde{V}^{\mathrm{cont}}+\widetilde{V}^{\mathrm{tens}}+\widetilde{V}^{\mathrm{so(v)}}+\widetilde{S}_{12}(r)+\widetilde{V}^{\mathrm{so(s)}},
\end{equation}
where {$\widetilde{G}_{12}$, $\widetilde{V}^{\mathrm{cont}}$, $\widetilde{V}^{\mathrm{tens}}$, $\widetilde{V}^{\mathrm{so(v)}}$, $\widetilde{S}_{12}(r)$, and $\widetilde{V}^{\mathrm{so(s)}}$} are the Coulomb potential term, the contact potential term, the tensor potential term,  the vector spin-orbit potential term, the screened confinement potential term  and the scalar spin-orbit interaction term, respectively.

The spin-independent potential terms of the  {nonrelativistic} potential read
\begin{equation}
\tilde{G}(r)=-\sum_k\frac{4\alpha_k}{3r}\left[\frac{2}{\sqrt{\pi}}\int_0^{\gamma_{k}r} e^{-x^2}dx\right],
\end{equation}
where $\alpha_k=(0.25, 0.15, 0.2)$ and $\gamma_k=(1/2{\,\text{GeV}},\sqrt{10}/2{\,\text{GeV}},\sqrt{1000}/2{\,\text{GeV}})$ for $k=1,2,3$ \cite{Godfrey:1985xj}, 
and
\begin{equation}
{S}(r)=\frac{b(1-e^{-\mu r})}{\mu}+c,
\end{equation}
where $\mu = 0.0779$ {GeV}  is the {screening parameter} given by our previous work \cite{Wang:2024lba} and listed in Table \ref{MGI}. This parameter characterizes the strength of the color screening effect, which is absent in the GI model and also represents an improvement of the MGI model over the GI model  \cite{Godfrey:1985xj}.
The color screening effect leads to a significant suppression of highly excited states' masses compared to the GI model \cite{Godfrey:1985xj}
. And $b$ is the confining  parameter, while $c$ is the vacuum constant. As shown in Table~\ref{MGI}, since the quark masses adopted in the MGI model of this work are all smaller than those in the GI model, the energy density $b$ of the interquark string must be increased to fit the mass spectrum, and the vacuum constant is also raised accordingly.
Here $b=0.222$ GeV$^2$ and $c=-0.228$ GeV, which are taken from Ref.~\cite{Wang:2024lba}.

The effective potential $\widetilde{V}^{\mathrm{eff}}$ {considers} relativistic effects, particularly in meson systems, which are embedded in two ways. First, a smearing function for a meson $q\bar{q}$ is introduced, which has the form
\begin{equation}
\label{smearing}
\rho_{ij} \left(\mathbf{r}-\mathbf{r'}\right)=\frac{\sigma_{ij}^3}{\pi ^{3/2}}e^{-\sigma_{ij}^2\left(\mathbf{r}-\mathbf{r'}\right)^2},
\end{equation}
with
\begin{align}
\label{smearing_sigma}
   \sigma_{ij}^2=\sigma_0^2\Bigg[\frac{1}{2}+\frac{1}{2}\left(\frac{4m_im_j}{(m_i+m_j)^2}\right)^4\Bigg]+
  s^2\left(\frac{2m_im_j}{m_i+m_j}\right)^2,
\end{align}
where $\sigma_0=1.791$ GeV is a universal parameter in Eq.~(\ref{smearing_sigma}), and $s=0.711$ is a parameter related to heavy quarkonium masses. Both the above two parameter values shown in Table \ref{MGI} are taken from Ref.~\cite{Wang:2024lba}.
The Coulomb term $\widetilde{G}_{12}(r)$ is defined as
\begin{equation}
\begin{split}
\widetilde{G}_{ij}(r)=&\int {\rm{d}}^3{\bf r}^\prime \rho_{ij}({\bf r}-{\bf r}^\prime)G(r^\prime)
=\sum\limits_k-\frac{4\alpha_k }{3r}{\rm erf}(\tau_{kij}r),
\end{split}
\end{equation}
where the values of $\tau_{kij}$ read
\begin{equation}
\tau_{kij}=\frac{1}{\sqrt{\frac{1}{\sigma_{ij}^2}+\frac{1}{\gamma_k^2}}}.
\end{equation}
The potential $\widetilde{S}_{12}(r)$ can be expressed as
\begin{eqnarray}
\widetilde{S}_{12}(r)&=& \int {\rm{d}}^3 {\bf r}^\prime
\rho_{12} ({\bf r}-{\bf r}^\prime)S(r^\prime)\nonumber\\
&=& \frac{b}{\mu r}\Bigg[r+e^{\frac{\mu^2}{4 \sigma^2}+\mu r}\frac{\mu+2r\sigma^2}{2\sigma^2}\Bigg(\frac{1}{\sqrt{\pi}}
\int_0^{\frac{\mu+2r\sigma^2}{2\sigma}}e^{-x^2}{\rm{d}}x-\frac{1}{2}\Bigg) \nonumber\\
&&-e^{\frac{\mu^2}{4 \sigma^2}-\mu r}\frac{\mu-2r\sigma^2}{2\sigma^2}\Bigg(\frac{1}{\sqrt{\pi}}
\int_0^{\frac{\mu-2r\sigma^2}{2\sigma}}e^{-x^2}{\rm{d}}x-\frac{1}{2}\Bigg)\Bigg]  \nonumber \\
&&+c. \nonumber\label{Eq:pot}
\end{eqnarray}
Second, due to relativistic effects, the general potential should depend on the mass of the interacting quarks. Momentum-dependent factors, which are unity in the nonrelativistic limit, are applied as
\begin{equation}
\tilde{G}_{12}(r)\to \tilde{G}_{12}=\left(1+\frac{\mathbf{p}^2}{E_1E_2}\right)^{1/2}\tilde{G}_{12}(r)\left(1 +\frac{\mathbf{p}^2}{E_1E_2}\right)^{1/2}.
\end{equation}
The semirelativistic  {corrections} of the spin-dependent terms are  written as
\begin{equation}
\label{vsoij}
  \tilde{V}^i_{\alpha \beta}(r)\to\tilde{V}^i_{\alpha \beta}= \left(\frac{m_\alpha m_\beta}{E_\alpha E_\beta}\right)^{1/2+\epsilon_i} \tilde{V}^i_{\alpha \beta}(r)\left(\frac{m_\alpha m_\beta}{E_\alpha E_\beta}\right)^{1/2+\epsilon_i},
\end{equation}
\begin{table}[htbp]
\renewcommand{\arraystretch}{1.5}
\caption{The value of $\gamma$ in the QPC model and the parameters of the MGI model, which are all taken from Ref. \cite{Wang:2024lba}. The parameter $\gamma$ characterizes the strength of quark pair creation in the QPC model. $m_{u(d)}$ and $m_s$ are the masses of the $u(d)$ and $s$ quarks (or antiquarks), $\mu$  is the {screening parameter}, $b$ is the confining  parameter, while $c$ is the vacuum constant. $\sigma_0$ is a universal parameter in Eq.~(\ref{smearing_sigma}), and $s$ is a parameter related to heavy quarkonium masses. $\epsilon_c$, $\epsilon_t$, $\epsilon_{\rm so(v)}$, and $\epsilon_{\rm so(s)}$ represent the relativistic corrections to the potential terms  $\widetilde{V}^{\mathrm{cont}}$, $\widetilde{V}^{\mathrm{tens}}$, $\widetilde{V}^{\mathrm{so(v)}}$, and $\widetilde{V}^{\mathrm{so(s)}}$, respectively.\label{MGI}}
\begin{center}
\begin{tabular}{cccc}
\hline\hline
Parameter &  value &Parameter &  value  \\
 \midrule[0.7pt]
$\gamma$  &10.16          &{$\sigma_0$ (GeV)}   &{1.791}\\
$m_{u(d)}$(GeV)    &0.162    &{$s$ }          &{0.711}\\
$m_s$ (GeV)       &0.377    &$\mu$ (GeV)          &0.0779 \\
$b$ (GeV$^2$)     &0.222    &$c$ (GeV)            &$-0.228$\\
$\epsilon_c$      &-0.137   &$\epsilon_{so(v)}$     &0.0550\\
$\epsilon_{so(s)}$  &0.366    &$\epsilon_t$         &0.493\\
  \hline\hline
\end{tabular}
\end{center}
\end{table}
where $\tilde{V}^i_{\alpha \beta}(r)$  {denote} the contact term, the tensor term, the vector and the scalar spin-orbit terms, and $\epsilon_i=\epsilon_c$, $\epsilon_t$, $\epsilon_{\rm so(v)}$, and $\epsilon_{\rm so(s)}$ impacts the potentials $\widetilde{V}^{\mathrm{cont}}$, $\widetilde{V}^{\mathrm{tens}}$, $\widetilde{V}^{\mathrm{so(v)}}$, and $\widetilde{V}^{\mathrm{so(s)}}$, respectively \cite{Wang:2021abg}. 
These relativistic correction parameters are larger than their counterparts in the GI model, which may also lead to more significant differences in the wave functions for meson masses involving spin-spin and spin-orbit couplings.
They have the following forms
\begin{equation}\label{Vcont}
\begin{split}
\widetilde{V}^{\mathrm{cont}}=\frac{2{\bf S}_1\cdot{\bf S}_2}{3m_1m_2}\nabla^2\widetilde{G}_{12}^c,
\end{split}
\end{equation}
\begin{equation}\label{Vtens}
\begin{split}
\widetilde{V}^{\mathrm{tens}}=&-\left(\frac{3{\bf S}_1\cdot{\bf r}{\bf S}_2\cdot{\bf r}/r^2-{\bf S}_1\cdot{\bf S}_2}{3m_1m_2}\right)\left(\frac{\partial^2}{\partial r^2}-\frac{1}{r}{\frac{\partial}{\partial r}}\right)\widetilde{G}_{12}^t,
\end{split}
\end{equation}
\begin{equation}\label{Vsov}
\begin{split}
\widetilde{V}^{\mathrm{so(v)}}=&\frac{{\bf S}_1\cdot {\bf L}}{2m_1^2}\frac{1}{r}\frac{\partial\widetilde{G}_{11}^{\rm so(v)}}{\partial r}+\frac{{\bf S}_2\cdot {\bf L}}{2m_2^2}\frac{1}{r}\frac{\partial\widetilde{G}_{22}^{\rm so(v)}}{\partial r}
\\
 &
+\frac{({\bf S}_1+{\bf S}_2)\cdot {\bf L}}{m_1m_2}\frac{1}{r}\frac{\partial\widetilde{G}_{12}^{\rm so(v)}}{\partial r},\\
\end{split}
\end{equation}

\begin{equation}\label{Vsos}
\begin{split}
\widetilde{V}^{\mathrm{so(s)}}=&-\frac{{\bf S}_1\cdot {\bf L}}{2m_1^2}\frac{1}{r}\frac{\partial\widetilde{S}_{11}^{\rm so(s)}}{\partial r}-\frac{{\bf S}_2\cdot {\bf L}}{2m_2^2}\frac{1}{r}\frac{\partial\widetilde{S}_{22}^{\rm so(s)}}{\partial r}.\\
\end{split}
\end{equation}

The Hamiltonian matrix in Eq.~(\ref{hh}) is diagonalized using the simple harmonic oscillator (SHO) basis, whose functional form is given by
\begin{equation}
 \begin{split}
	\psi_{nLM_L}^{SHO}(\mathbf{p})=R_{nL}^{SHO}(p, \beta)Y_{LM_L}(\Omega_p),
\end{split}
\end{equation}
with
\begin{align} \label{1.3}
R_{nL}^{SHO}(p,\beta)=\frac{(-1)^{(n-1)}(-i)^L}{\beta^{3/2}}N_{nL}e^{-\frac{p^2}{2\beta^2}}{(\frac{p}{\beta})}^{L} \times L_{n-1}^{L+1/2}(\frac{p^2}{ \beta ^2}),
\end{align}
\begin{align} \label{1.4}
N_{nL}=\sqrt{\frac{2(n-1)!}{\Gamma(n+L+1/2)}},
\end{align}
where $Y_{LM_L}(\Omega)$ is a spherical harmonic function, $L_{n-1}^{L+1/2}(x)$ is  the associated Laguerre polynomial and $\Gamma(n+L+1/2)$ is the gamma function. {Then the mass spectra of $0^+$ light strangemesonfamily can be obtained, which are listed in Table \ref{mass}.}
The spatial wave function of the mesons can be expressed as
\begin{align} \label{expand}
R_{nL}(p)=\sum_{n=1}^{n_{\rm max}}C_{n}{R}_{nL}^{\rm SHO}(p,\beta).
\end{align}
Then the SHO wave function depends only on a single parameter: the harmonic oscillator scale $\beta$. If a wave function is expanded in a complete set of SHO basis functions, then in principle, the solution of the potential model should be independent of the oscillator parameter $\beta$. However, in practical calculations, the number of basis functions used in the expansion is limited, which necessitates choosing an appropriate value for the parameter $\beta$. 
An effective method for selecting the parameter $\beta$ is to choose the value that minimizes the eigenvalue $E_{nL}$, i.e., ($\frac{\partial E_{nL}}{
\partial\beta_{i}}=0$, {$\frac{\partial^2 {E_{nL}}}{
\partial\beta_{i}^2}$}$>0$), where $i$ denotes different light mesons.

In this work, we set $n_{max}=21$. For the mesons  $K_0^*(2P)(K_0^*(1950))$,  $K_0^*(4P)$, $K_0^*(5P)(\kappa(2600))$, and $K_0^*(6P)$, the corresponding $\beta$ values take  0.475 GeV, 0.328 GeV, 0.301 GeV, and 0.258 GeV, respectively. The $\beta$ values for the other mesons mentioned in this work are  {listed} in Table \ref{beta} and the expansion coefficients $C_n$ of  $K_0^*(2P)$, $K_0^*(4P)$, $K_0^*(5P)$, $K_0^*(6P)$, $K^*$ and $\phi$ are {listed} in Table \ref{wavec}.
\par
\begin{figure}[htbp]
\includegraphics[scale=0.7]{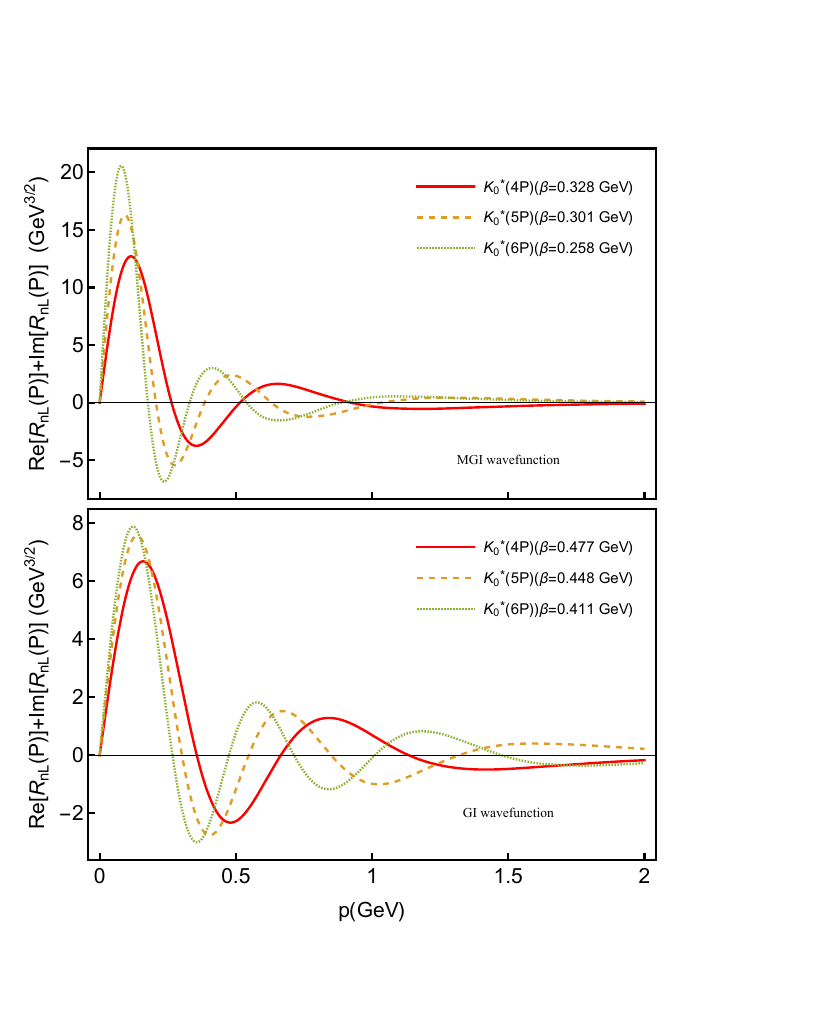}
\caption{The model dependence of the radial wave functions' shapes of $K_0^*(4P)$, $K_0^*(5P)$, and $K_0^*(6P)$ derived from (a) the GI potential model (GI wave function); and (b) the MGI potential model (MGI wave function). “Re" and “Im" denote the real part and the imaginary part of the wave functions, respectively.\label{wave1fig}}
\end{figure}

We give the model dependence of the radial wave functions' shapes of $K_0^*(4P)$, $K_0^*(5P)$, and $K_0^*(6P)$ obtained with the MGI model (MGI wave function) and the GI model (GI wave function) \cite{Godfrey:1985xj}  in Fig.~\ref{wave1fig}.  The choice of potential model significantly affects the radial spatial wave functions (such as the positions of nodes and the maximum values). 

\begin{figure}[htbp]
\includegraphics[scale=0.7]{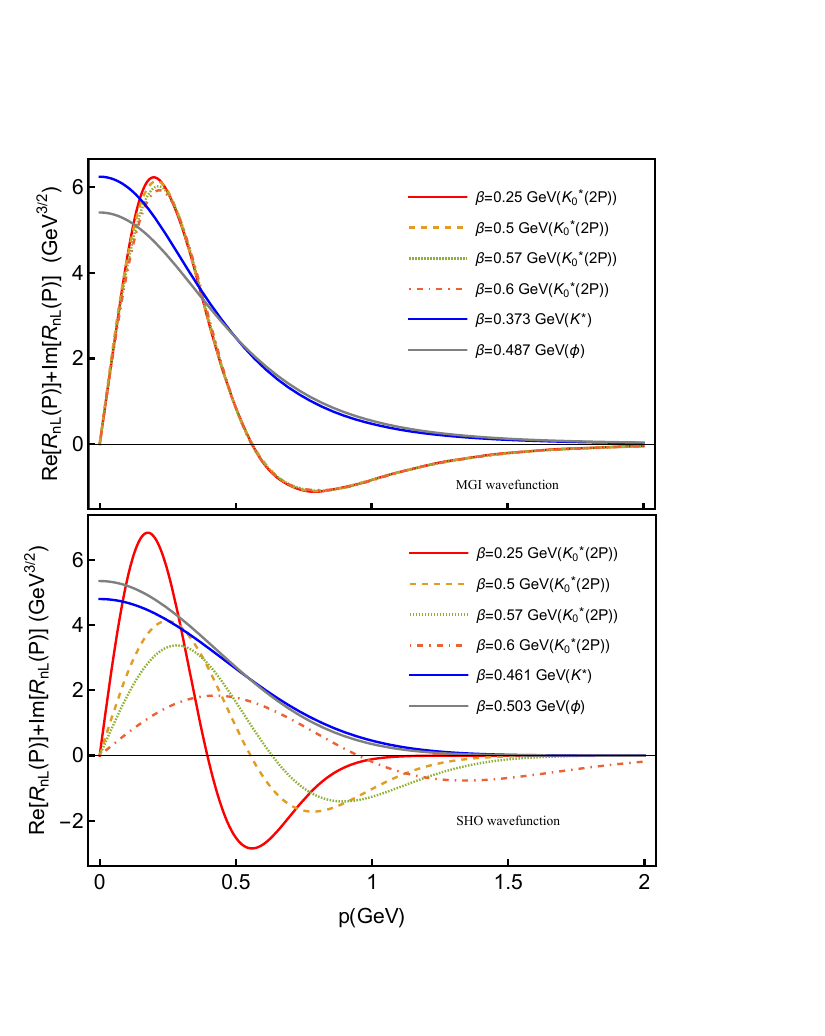}
\caption{
(a) The $\beta$ dependence of the radial wave functions' shapes of $K_0^*(2P)$, and the radial wave functions' shapes of $K^*$ and $\phi$ derived from  the MGI model (MGI wave function); 
(b)  The $\beta$ dependence of the radial wave functions' shapes of $K_0^*(2P)$, and the radial wave functions' shapes of $K^*$ and $\phi$ from the SHO wave function directly. The $\beta$ values for the SHO wave functions of  $K^*$ and $\phi$, referred to as effect $\beta$, are calculated based on the MGI wave functions \cite{Wang:2021hho}. “Re" and “Im" denote the real part and the imaginary part of the wave functions, respectively. \label{wave2fig}}
\end{figure}

In Fig.~\ref{wave2fig}, we show the $\beta$ dependence of MGI wave function  of $K_0^*(2P)$ and SHO wave function. We can conclude that the $\beta$ value (in the range of 0.25 GeV-0.6 GeV) has very little influence on the shape of MGI wave function as shown in Fig.~\ref{wave2fig}(a). The  wave functions of $K^*$ and $\phi$ have no node as the ground states.
However, it can be seen from Fig.~\ref{wave2fig}(b) that for the SHO wave function of $K_0^*(2P)$, the node positions are sensitive to the value of $\beta$.

In the QPC model, the decay width has a strong dependence on the shape of the wave function, because of the node effect \cite{Duan:2020tsx}. We will discuss this point in the next subsection.

The mass and the spatial wave function of the meson can be obtained by diagonalizing Hamiltonian matrix in Eq.~(\ref{hh}) and applied to {the calculation of} the strong decay processes. 

\subsection{The QPC model}
\begin{figure*}[htbp]
\includegraphics[scale=0.7]{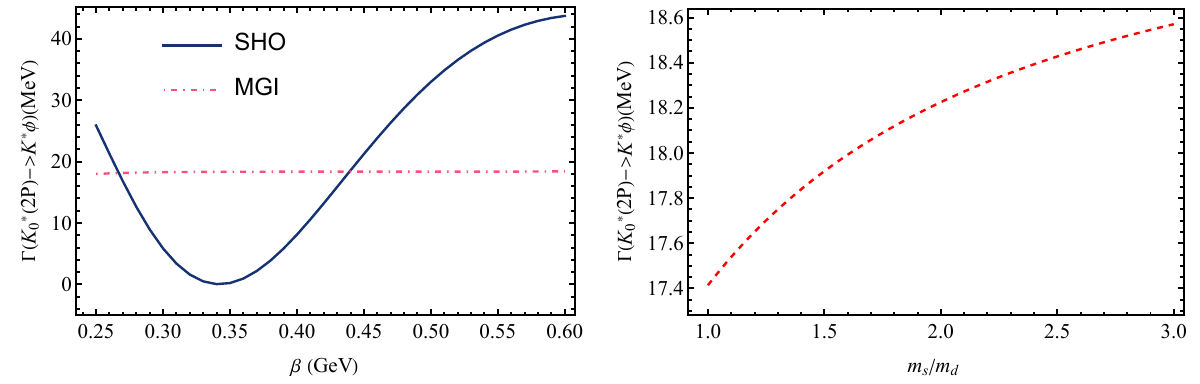}
\caption{(a) The $m_s/m_d$ ratio dependence of the decay width of $K_0^*(2P)\to K^*\phi$ obtained with the MGI wave function.  (b) Decay width of $K_0^*(2P)\to K^*\phi$ as a function of the harmonic oscillator scale $\beta$. The solid line and the dashed line represent the widths depending on the $\beta$ values when we take the SHO wave function and the MGI wave function, respectively, as taken from Fig.~\ref{wave2fig}.
\label{widthbeta}}
\end{figure*}

The QPC model (or $^3P_0$ model) was initially {proposed} by Micu \cite{Micu:1968mk} and further developed by the Orsay group \cite{LeYaouanc:1972ae, LeYaouanc:1973xz, LeYaouanc:1974mr, LeYaouanc:1977gm, LeYaouanc:1977ux}. This model is widely applied to calculate the OZI-allowed two-body strong decays of mesons \cite{vanBeveren:1982qb, Titov:1995si, Ackleh:1996yt, Blundell:1996as, Bonnaz:2001aj, Zhou:2004mw, Lu:2006ry, Zhang:2006yj, Luo:2009wu, Sun:2009tg, Liu:2009fe, Sun:2010pg, Rijken:2010zza, Ye:2012gu, Wang:2012wa, He:2013ttg, Sun:2013qca,  Pang:2018gcn, Wang:2022juf, Wang:2022xxi, Li:2022khh, Li:2022bre, Wang:2020due, Pang:2017dlw, Wang:2024yvo, Wang:2021abg, feng:2021igh, Feng:2022hwq}.
The $\mathcal{T}$ operator describes a quark-antiquark pair (denoted by indices 3 and 4) creation from vacuum, which reads
$J^{PC}=0^{++}$.
For the process $A\to B+C$, $\mathcal{T}$ can be written as 
{\begin{align}\label{gamma}
\mathcal{T} = & -3\gamma \sum_{m}\langle 1m;1~-m|00\rangle\int d \mathbf{p}_3d\mathbf{p}_4\delta ^3 (\mathbf{p}_3+\mathbf{p}_4) \nonumber \\
 & ~\times \mathcal{Y}_{1m}\left(\frac{\textbf{p}_3-\mathbf{p}_4}{2}\right)\chi _{1,-m}^{34}\phi _{0}^{34}
\left(\omega_{0}^{34}\right)_{ij}b_{3i}^{\dag}(\mathbf{p}_3)d_{4j}^{\dag}(\mathbf{p}_4),
\end{align}
where the parameter $\gamma$ in QPC model depicts the strength of $q\bar{q}$ creation from the vacuum. In this work, we adopt the value $\gamma  = 10.16$~\cite{Wang:2024lba}.
The decay width is proportional to {$\gamma ^{2}$}, whereas the branching ratio typically shows no dependence on this parameter. $\mathcal{Y}_l^m(\bf{p})\equiv$ $p^lY_l^m(\theta_p,\phi_p)$ is the solid harmonics. $\chi$, $\phi$, and $\omega$ denote the spin, flavor, and color wave functions respectively. $\mathbf{p}_3$ and $\mathbf{p}_4$ denote the three-momenta of the quark pair {created} in vacuum. Subindices $i$ and $j$ are the color index of the $q\bar{q}$ pair. The amplitude $\mathcal{M}^{{M}_{J_{A}}M_{J_{B}}M_{J_{C}}}$ has the form
\begin{eqnarray}
\langle BC|\mathcal{T}|A \rangle = \delta ^3(\mathbf{P}_B+\mathbf{P}_C)\mathcal{M}^{{M}_{J_{A}}M_{J_{B}}M_{J_{C}}},\end{eqnarray}
where $\bf{P_B}$  and $\bf{P_C}$ are the three-momentums of the meson B and the meson C in the rest frame of the meson A,  {${M}_{J_{X}}$ (with $X=A,B,C$) denotes the magnetic quantum number of the the corresponding meson.}
Finally, the general form of the decay width can be expressed as
\begin{eqnarray}
\Gamma&=&\frac{\pi}{4} \frac{|\mathbf{P}|}{m_A^2}\sum_{J,L}|\mathcal{M}^{JL}(\mathbf{P})|^2,
\end{eqnarray}
where $m_{A}$ is the mass of the initial state $A$, $\bf{J}=\bf{J_B}+\bf{J_C}$, $L$ is the relative orbital angular momentum between the meson B and the meson C, and
$\bf{P}=\bf{P_B}$. The two decay amplitudes can be related by the Jacob-Wick formula \cite{Jacob:1959at} }
\begin{equation}
\begin{aligned}
\mathcal{M}^{JL}(\mathbf{P}) = &\frac{\sqrt{4\pi(2L+1)}}{2J_A+1}\sum_{M_{J_B}M_{J_C}}\langle L0;JM_{J_A}|J_AM_{J_A}\rangle \\
    &\times \langle J_BM_{J_B};J_CM_{J_C}|{J_A}M_{J_A}\rangle \mathcal{M}^{M_{J_{A}}M_{J_B}M_{J_C}},
    \end{aligned}	
\end{equation}

in which
\begin{equation}
\begin{aligned}
 &\mathcal{M}^{M_{J_A} M_{J_B} M_{J_C}}\\
 &=\gamma \sum_{\substack{M_{L_A}, M_{S_A}, M_{L_B},\\ M_{S_B}
M_{L_C, M_S}, m}}\left\langle L_A M_{L_A} S_A M_{S_A} \mid J_A M_{J_A}\right\rangle \\
& \times\left\langle L_B M_{L_B} S_B M_{S_B} \mid J_B M_{J_B}\right\rangle\left\langle L_C M_{L_C} S_C M_{S_C} \mid J_C M_{J_C}\right\rangle \\
& \times\langle 1 m 1-m \mid 00\rangle\left\langle\chi_{S_B M_{S_B}}^{14} \chi_{S_C M_{S_C}}^{32} \mid \chi_{S_A M_{S_A}}^{12} \chi_{1-m}^{34}\right\rangle \\
& \times\left[\left\langle\phi_B^{14} \phi_C^{32} \mid \phi_A^{12} \phi_0^{34}\right\rangle I\left(\mathbf{P}, m_1, m_2, m_3\right)\right.+ \\
& \left.(-1)^{1+S_A+S_B+S_C+L_C}\left\langle\phi_B^{32} \phi_C^{14} \mid \phi_A^{12} \phi_0^{34}\right\rangle I\left(-\mathbf{P}, m_2, m_1, m_3\right)\right], \\
\end{aligned}
\end{equation}
with
\begin{equation}
\begin{aligned}
&I\left(\mathbf{P}, m_1, m_2, m_3\right)\\
&=\int d^3 \mathbf{k} \psi_B^*(\mathbf{k}+U \mathbf{P}) \psi_C^*(\mathbf{k}+V\mathbf{P} ) \psi_A(\mathbf{k}-\mathbf{P}) \mathcal{Y}_1^m(\mathbf{k}),
\end{aligned}
\end{equation}
here 
\begin{equation*}
\begin{aligned}
U=\frac{m_3}{m_1+m_3}, V=\frac{m_3}{m_2+m_3},
\end{aligned}
\end{equation*}
$m_1$ and $m_2$ are the masses of the quark and the antiquark in meson A, respectively. For the $K$ meson, we take $m_1=0.377$ GeV, $m_2=0.162$ GeV. $m_3$ denotes the mass of the quark (or the antiquark) created from the vacuum, and is set to 0.162 GeV for $u \bar u$ or $d\bar d$, and 0.377 GeV for $s \bar s$. To investigate the quark mass effect on decays, we calculate the decay width of $K_0^*(2P)\to K^*\phi$ as a function of the ratio $m_s/m_d$ as shown in Fig.~\ref{widthbeta}({a}) and find that the decay width is not sensitive to this mass ratio in the range of $1\sim3$.

When we calculate the strong decays of {the} $K_0^*$ family using QPC model in this work, the spatial wave functions of mesons are obtained using the MGI model (MGI wave function), as introduced in the previous subsection. 

{As previously discussed}, the SHO wave function is characterized by a single scale parameter, $\beta$. According to Ref. \cite{Duan:2020tsx}, the node structure of the radial wave function can influence the strong decay width of {the} process $A\to BC$, known as the node effect. Specifically, the spatial wave function $R_{nL}(p)$ for the $n$-th radial excitation contains $(n-1)$ nodes. In Fig.~\ref{widthbeta}({b}), we give the line shape of the decay width of $K_0^*(2P)\to K^*\phi$ depending on the $\beta$ values. 
{When the SHO wave function is adopted, the decay width exhibits strong $\beta$ dependence, which is consistent with the conclusions in Ref.~\cite{Duan:2020tsx}, as shown in Fig.~\ref{widthbeta}({b}).}
In contrast, when the MGI wave function is used, the width shows virtually no $\beta$ sensitivity. When employing different potential models (such as the GI model) to derive wave functions, decay widths will also vary even with identical values for other parameters (such as quark masses and $\gamma$).

\par

Among the final states, the mixing scheme for natural states ($L=J$) of strange mesons can be expressed as
\begin{equation}\label{anglek1}
\left( \begin{matrix}
	|K(nL)\rangle \\
	|K^\prime( nL)\rangle \\
\end{matrix}\right) =
\left( \begin{matrix}
	\textrm{$\cos\theta_{nL}$} & \textrm{$\sin\theta_{nL}$} \\
	\textrm{$-\sin\theta_{nL}$} & \textrm{$\cos\theta_{nL}$} \\
\end{matrix}\right)
\left( \begin{matrix}
	|K(n^1L_L)\rangle \\
	|K(n^3L_L)\rangle \\
\end{matrix}\right),
\end{equation}
where $\theta_{nL}$ is known as the mixing angle between the $K(n^1L_L)$ and $K(n^3L_L)$ states. In this work, the masses of ${K(nL)}$ and ${K^\prime (nL)}$, the mixing angle $\theta_{nL}$ for the missing states are also calculated via the MGI model \cite{Wang:2024yvo} and listed in Table \ref{beta}. 

The flavor wave functions of the isoscalar mesons may have the mixing form
\begin{equation}\label{mixingns}
\left( \begin{matrix}
	X \\
	X^\prime \\
\end{matrix}\right) =
\left( \begin{matrix}
	\textrm{$\cos\phi_{x}$} & \textrm{$\sin\phi_{x}$} \\
	\textrm{$-\sin\phi_{x}$} & \textrm{$\cos\phi_{x}$} \\
\end{matrix}\right)
\left( \begin{matrix}
	|n\bar{n}\rangle \\
	|s\bar{s}\rangle \\
\end{matrix}\right),
\end{equation}
where $X$ and $X^\prime$ are two isoscalar mesons (such as $\eta$ and $\eta^\prime$), $\phi_x$ is the mixing angle in the quark flavor scheme, and the flavor $n\bar{n}=(u\bar{u}+d\bar{d})/\sqrt{2}$. We also list the flavor mixing information of the isoscalar mesons appearing in this paper in Table \ref{beta}.

\section{Mass spectra and {two-body} strong decay of the highly excited states of {$K_0^{*}$}  family} \label{sec3}

In this section, we discuss the mass spectra and {two-body} strong decay information of the highly excited {states} of $0^{+}$ strange meson family. We start with the mass spectra.
\subsection{Mass spectra analysis} 

We employ the MGI model to calculate the mass spectra of $0^{+}$ light strange meson family. The parameters of the MGI model listed in Table \ref{MGI} are obtained by fitting the experimental masses of light mesons in our previous work \cite{Wang:2024lba}. These parameters are universal and can also be used in the study of masses of other light mesons. Reference \cite{Ebert:2009ub} calculated the mass {spectra} of the light mesons within the framework of the relativistic quark model based on the quasipotential approach. This quasipotential model has {a} similar string tension {with  the GI model}. Experimental values and theoretical results from {the} GI model, {the} MGI model (this work) and Ref. \cite{Ebert:2009ub} are shown in Table \ref{mass}.
\begin{table*}[htb] 
 \small
 \renewcommand{\arraystretch}{1.5}
\centering
\caption{The mass spectra of $n^3P_0$ states. The unit {of mass} is MeV. \label{mass}}
\[\begin{array}{ccccccccccccccc} 
\hline
\hline
&  & n^{2s+1}L_J &\text{State}  & \text{This work}& \text{GI}$~ \cite{Godfrey:1985xj}$&\text{Ebert}$~\cite{Ebert:2009ub}$&\text{Exp. }$\cite{ParticleDataGroup:2024cfk}$
\\\midrule[0.7pt]
&  & 1^3P_0  &K_0^*(1430)    &1284  &1233&1362 & 1425\pm50~$\cite{ParticleDataGroup:2024cfk}$     \\
&  & 2^3P_0  &K_0^*(1950)    &1829 &1890&1791&  1957\pm 14~$\cite{ParticleDataGroup:2024cfk}$   \\
&  & 3^3P_0  &K_0^*(2130)    &2185 &2356&2160  &2128\pm31\pm9~$\cite{BaBar:2021fkz}$                 \\
&  & 4^3P_0  &K_0^*(4P)      &2451 &2749&-&-                  \\
&  & 5^3P_0  &K_0^*(5P)      &2659 &3092&-&  2662\pm59\pm201   ~$\cite{LHCb:2023evz}$             \\
&  & 6^3P_0  &K_0^*(6P)      &2832 &3404&-& -                \\
 \hline
 \hline
\end{array}\] 
\end{table*}

According to {Table~}\ref{mass}, for the $1^3P_0$ state, our theoretical value {of the mass} ({1284} MeV) remains $\sim$140 MeV below experimental value $(1425 \pm 50)$ MeV. {The} GI model \cite{Godfrey:1985xj} predicts 1233 MeV ($\sim$192 MeV deficit), Ebert's result of 1362 MeV is still $\sim$60 MeV {lower}  \cite{Ebert:2009ub}.
We will now focus on discussing the spectra of the highly excited states.
Our predictions of $3^3P_0$  ({2185} MeV)  overlap with experimental error bars of $K_0^*(2130)$ (2128$\pm$40 MeV), {the} GI model shows $\sim$230 MeV deviation, and Ebert's {result of 2160 MeV is} also consistent with the experimental value. This result shows that introducing the color screening effect into the GI model can significantly improve the predictive power for the high-excited-state meson spectrum.

For {the} $K_0^*(4P)$ state, our theoretical mass of 2451 MeV is 300 MeV lower than the value given by the GI model. We predict that $K_0^*(5P)$ has  {a} mass of 2658 MeV, which is in excellent agreement with the experimental central value.

We also calculate the mass of $K_0^*(6P)$ {to be} 2832 MeV, approaching the mass range of charmed mesons. 
\par
Through the analysis of the mass spectrum, we conclude that the newly observed $\kappa(2600)$ meson in experiments is a good  candidate for the $K_0^*(5P)$ state. Subsequently, we will further validate this conclusion by investigating its decay width and predicting the two-body strong decay properties of  $K_0^*(4P)$ and $K_0^*(6P)$.

\subsection{Strong decay analysis} 
In this section, we discuss the two-body strong decay of the highly excited states {in the} $K_0^*$ family.
At first, we test our model with the ground state $K_0^*$. We obtain two decay channels of $K_0^*(1430)$, $K\pi$ and $\eta K$ as shown in Table \ref{1P}, in which the decay channels with {branching} radios (BR) less than 0.2\% are omitted. 

The $K\pi$ channel accounts for 95.5\% of total width ($\Gamma_{K\pi} = 252$~MeV), consistent with experimental branching ratio $0.93 \pm 0.1$. {The predicted  branching ratio of $\frac{\Gamma_{\eta K}}{\Gamma_{Total}}= 4.2\%$ is roughly consistent with the experimental value $0.086^{+0.037}_{-0.034}$.} The calculated total width $\Gamma_{\text{total}} = 264$~MeV {shows} excellent agreement with {the} experimental data $270 \pm 80$~MeV. 
These results show that the QPC model can work well for $K_0^*(1430)$, and we expect it to also be suitable for  the highly excited states of {the} $K_0^*$ family.

\begin{table*}[htb]
\small
\renewcommand{\arraystretch}{1.2}
\centering
\caption{
The $\beta$ values in Eq.~(\ref{expand}), the masses of the missing states, as well as some flavor mixing angles or spin mixing angles for each state of the light mesons mentioned in this work. 
The particles  $\omega$, $\rho$, $\eta^\prime$, $\phi$, $b_1$, $h_1$, $h_1^\prime$,  $a_0$, $a_1$, $a_2$, $f_1$,  $K^*$, $K_1$, ${K_1}^\prime$,  and $K_2^*$ correspond to the mesons  $\omega(783)$, $\rho(770)$, $\eta^\prime(958)$, $\phi(1020)$, $b_1(1235)$, $h_1(1170)$, $h_1(1380)$, $a_0(1450)$, $a_1(1260)$, $a_2(1320)$, $f_1(1285)$,   $K^*(892)$, $K_1(1270)$, ${K_1}(1400)$, and  $K_2^*(1430)$, respectively. $\eta(4S)$, $K_1(2P)$, $K_1(3P)$, $\pi(4S)$,  and so on, represent the states that exist theoretically but have not yet been observed experimentally in Table \ref{nP}, and their masses are calculated via the MGI model. 
$\beta\text{(Component)}$ represents  
the $\beta$ value of the component of the meson.
{For example, } ${[}646(n\bar n*\text{cos}(-39.3^\circ)), 781 (s\bar s*\text{sin} (-39.3^\circ)){]}$ represents the flavor wave function of $\eta${, which} is $(n\bar n*\text{cos}(-39.3^\circ))+(s\bar s*\text{sin} (-39.3^\circ))$, the  $\beta$ value of the spatial wave function of $n\bar n$ ($(u\bar{u}+d\bar{d})/\sqrt{2}$) component  is 646 MeV, and 
the  $\beta$ value of the spatial wave function of $s\bar s$ component  is  781 MeV.
The units of mass and $\beta$ are MeV.  
\label{beta}}
\vspace{-15pt}
 \[\begin{array}{cccccccc}
  \hline
 \hline
 \text{State(Mass)} & {\beta} & \text{State(Mass)} & \beta\text{(Component)} \text{State(Mass)} & {\beta} & \text{State(Mass)} & \beta\text{(Component)} \\
  \hline
 K^* &373& \eta ~$\cite{Feldmann:1999uf}$ & \begin{array}{c} 646(n\bar n*\text{cos}(-39.3^\circ)) \\ 781 (s\bar s*\text{sin} (-39.3^\circ))\end{array} &
 
 K_0^*\text{(1430)} & 556 &
\eta^\prime~$\cite{Feldmann:1999uf}$ & \begin{array}{c} 646(-n\bar n*\text{sin}(-39.3^\circ)) \\781 (s\bar s*\text{cos} (-39.3^\circ))\end{array} \\

K_0^*\text{(2130)} & 407 &
 \eta^\prime\text{(1475)}~ $\cite{Yu:2011ta}$ & \begin{array}{c} 507(-n\bar n*\text{sin}(-39.3^\circ)) \\580 (s\bar s*\text{cos} (-39.3^\circ))\end{array} &
 
a_0\text{}& 541 &\text{$\eta $(4S)}(2119) &420(n\bar n)\\

 a_0(1740) &475&  K_2\text{(1770)} ~$\cite{Pang:2017dlw} $& \begin{array}{c} 458(1^1D_2 \text{cos}(-39^\circ)) \\ 
 427(1^3D_2\text{sin}(-39^\circ)) \end{array}&

 \rho_3\text{(1690)} & 279  & K_2^\prime\text{(1820)}~$\cite{Pang:2017dlw}$& \begin{array}{c} 458(-1^1D_2 \text{sin}(-39^\circ)) \\ 
 427(1^3D_2\text{cos}(-39^\circ)) \end{array}\\

 \pi  & 646& K_1~$\cite{Cheng:2013cwa} $& \begin{array}{c} 566(1^1P_1 \text{cos}(-34^\circ)) \\ 
 537(1^3P_1\text{sin}(-34^\circ)) \end{array}&

  \text{$\pi $(1300)} & 507 & K_1\text{(2P)}(1916)& \begin{array}{c} 434(2^1P_1 \text{cos}(-42^\circ)) \\ 
 447(2^3P_1\text{sin}(-42^\circ)) \end{array}\\

\text{$\pi $(1800)} & 484  &  K_1\text{(3P)}(2260) & \begin{array}{c} 410(3^1P_1 \text{cos}(-42^\circ)) \\ 
 382(3^3P_1\text{sin}(-42^\circ)) \end{array}&

\text{$\pi $(4S)}(2119) & 363 & K_1\text{(4P)}(2517) & \begin{array}{c} 336(4^1P_1 \text{cos}(-43^\circ)) \\ 
 328(4^3P_1\text{sin}(-43^\circ)) \end{array}\\

 K & 632 & K_1^\prime~$\cite{Cheng:2013cwa} $& \begin{array}{c} 566(-1^1P_1 \text{sin}(-34^\circ)) \\ 
 537(1^3P_1\text{cos}(-34^\circ)) \end{array}&

 {K(1460)} & 524& K_1^\prime\text{(2P)}(1760)& \begin{array}{c} 434(-2^1P_1 \text{sin}(-42^\circ)) \\ 
 447(2^3P_1\text{cos}(-42^\circ)) \end{array}\\

  {K(3S)}(1917) & 476  &  K_1^\prime\text{(3P)}(2111) & \begin{array}{c} 410(-3^1P_1 \text{sin}(-42^\circ)) \\ 
 382(3^3P_1\text{cos}(-42^\circ)) \end{array}&
 
   {K(4S)} (2261)& 380 & K_1^\prime\text{(4P)}(2388) & \begin{array}{c} 336(-4^1P_1 \text{sin}(-43^\circ)) \\ 
 328(4^3P_1\text{cos}(-43^\circ)) \end{array}\\

   {K(5S)} (2517)& 343 & K_1^\prime\text{(5P)}(2608) & \begin{array}{c} 306(-5^1P_1 \text{sin}(-44^\circ)) \\ 
 295(5^3P_1\text{cos}(-44^\circ)) \end{array}&

 \pi _2\text{(1880)}~$\cite{Ebert:2009ub}$ & 336 &  K_2\text{(2D)}(2165)& \begin{array}{c} 365(2^1D_2 \text{cos}(-44^\circ)) \\ 
 373(2^3D_2\text{sin}(-44^\circ)) \end{array}\\

 \pi _2\text{(3D)}(2273) &  317    &K_2^\prime\text{(2D)}(2074)& \begin{array}{c} 365(-2^1D_2 \text{sin}(-44^\circ)) \\ 
 373(2^3D_2\text{cos}(-44^\circ)) \end{array}&

\rho_2(1D)(1644)&411&K_2^\prime\text{(3D)}(2355)& \begin{array}{c} 339(-3^1D_2 \text{sin}(-44^\circ)) \\ 
 338(3^3D_2\text{cos}(-44^\circ)) \end{array}\\

 \rho _2\text{(2D)}(2001)& 343&  K_2\text{(4D)}(2580)& \begin{array}{c} 293(4^1D_2 \text{cos}(-45^\circ)) \\ 
 290(4^3D_2\text{sin}(-45^\circ)) \end{array}&

   \rho& 303  &  K_3\text{(1F)}(2028))& \begin{array}{c} 368(1^1F_3 \text{cos}(-44^\circ)) \\ 
 410(1^3F_3\text{sin}(-44^\circ)) \end{array}\\

 K_2^*\text{(1980)} & 395 &  K_3\text{(3F)}(2545))& \begin{array}{c} 292(1^1F_3 \text{cos}(-45^\circ)) \\ 
 301(1^3F_3\text{sin}(-45^\circ)) \end{array}&

 a_3\text{(2030)}~$\cite{Ebert:2009ub}$ & 311 &  K_3^\prime\text{(2F)}(2312))& \begin{array}{c} 320(-2^1F_3 \text{sin}(-44^\circ)) \\ 
 316(2^3F_3\text{cos}(-44^\circ)) \end{array}\\

  \rho(1450) ~$\cite{He:2013ttg}$&477   &  K_3^\prime\text{(3F)}(2545))& \begin{array}{c} 292(-3^1F_3 \text{sin}(-45^\circ)) \\ 
 301(3^3F_3\text{cos}(-45^\circ)) \end{array}&

K_2^* &344 &  K_4\text{(1G)}(2274)& \begin{array}{c} 315(4^1D_2 \text{cos}(-45^\circ)) \\ 
 337(4^3D_2\text{sin}(-45^\circ)) \end{array}\\

  a_1 & 475& h_1~$\cite{Li:2005eq}$ & \begin{array}{c} 571(n\bar n*\text{cos}(4.4^\circ)) \\ 631 (-s\bar s*\text{sin} (4.4^\circ))\end{array} &
 
a_1\text{(1640)}~$\cite{Chen:2015iqa} $ & 416&
h_1^\prime~$\cite{Li:2005eq}$& \begin{array}{c} 571(n\bar n*\text{-sin}(4.4^\circ)) \\
 631 (s\bar s*\text{cos} (4.4^\circ))\end{array} \\

 a_1\text{(3P)}(2065) & 364& h_1\text{(1595)} ~$\cite{Wang:2024lba}$ & \begin{array}{c} 404(n\bar n*\text{cos}(4.4^\circ)) \\ 482 (-s\bar s*\text{sin} (4.4^\circ))\end{array} &
 
  b_1 & 571& h_1\text{(1965)} ~$\cite{Wang:2024lba}$ & \begin{array}{c} 398(n\bar n*\text{cos}(4.4^\circ)) \\ 445 (-s\bar s*\text{sin} (4.4^\circ))\end{array} \\

  b_1\text{(2P)}(1699) & 404& h_1\text{(2215)}~ $\cite{Wang:2024lba}$ & \begin{array}{c} 313(n\bar n*\text{cos}(4.4^\circ)) \\ 368 (-s\bar s*\text{sin} (4.4^\circ))\end{array} &

b_1\text{(1960)}~$\cite{Chen:2015iqa}$  & 398    & \omega& 303(n\bar n) \\

    b_1\text{(2240)}~$\cite{Chen:2015iqa}$ & 313    &  f_3\text{(2050)}~$\cite{Ebert:2009ub}$ & \begin{array}{c}  311(n\bar n)\end{array}&
    
  a_2 & 293& f_1 ~$\cite{Yang:2010ah}$ & \begin{array}{c}  475(n\bar n*\text{cos}(-15.8^\circ)) \\   588(s\bar s*\text{sin} (-15.8^\circ))\end{array} \\

   a_3(2275)~$\cite{Ebert:2009ub}$& 254 & f_1^\prime ~$\cite{Yang:2010ah}$ & \begin{array}{c}  475(-n\bar n*\text{sin}(-15^\circ)) \\   588(s\bar s*\text{cos} (-15^\circ))\end{array} &

     \rho(1700)~$\cite{He:2013ttg}$ &484  & f_1\text{(2P)} (1717)& \begin{array}{c}  395(n\bar n)\end{array} \\

    K^*(1680)&  487& f_1^\prime\text{(2P)}(1739)  & \begin{array}{c}  470(s\bar s)\end{array}&

K^*\text{(3S)} &  379& f_1\text{(3P)} (2065)& \begin{array}{c}  364(n\bar n)\end{array} \\

 K^*\text{(1410)} & 480  & f_2\text{(2P)} (1739)& \begin{array}{c}  373(n\bar n)\end{array} &

a_2{(1700)} ~$\cite{Pang:2014laa}$& 395  &\phi& 488(s\bar s)  \\

 \hline
 \hline
\end{array}\]
\end{table*}

\begin{table*}[htb]
\small
 \renewcommand{\arraystretch}{1.2}
\centering
\caption{The values of $C_n$ ($n=1\sim21$) to reproduce the numerical radial wave functions (obtained with the MGI model) of $K_0^*\text{(2P)}$,$K_0^*\text{(4P)}$, $K_0^*\text{(5P)}$, $ K_0^*\text{(6P)}$  and the final {channel} $K^*\phi$ in
 Fig.~\ref{wave1fig}, Fig.~\ref{wave2fig}, and Fig.~\ref{widthbeta}.
 \label{wavec}}
\[\begin{array}{ccccccc}
\hline\hline
C_n&K_0^*\text{(2P)}&K_0^*\text{(4P)} &K_0^*\text{(5P)} &K_0^*\text{(6P)}&K^*&\phi\\
\hline
 C_1 & 0.222449 & -0.0555655 & 0.0140115 & -0.00276401 & 0.972251 & 0.98974 \\
 C_2 & 0.75125 & -0.28638 & -0.119631 & -0.0653887 & 0.157035 & -0.00561464 \\
 C_3 & -0.353308 & 0.228128 & -0.360601 & 0.0791252 & 0.150245 & 0.135653 \\
 C_4 & 0.388185 & 0.614452 & 0.337153 & 0.502814 & 0.0582049 & 0.0117256 \\
 C_5 & -0.20643 & -0.211652 & 0.402642 & -0.156157 & 0.0492352 & 0.0379793 \\
 C_6 & 0.193626 & 0.486748 & -0.121086 & -0.323823 & 0.026903 & 0.00892844 \\
 C_7 & -0.107969 & -0.195027 & 0.47625 & 0.107893 & 0.0216017 & 0.0144072 \\
 C_8 & 0.0999782 & 0.303553 & -0.205859 & -0.476104 & 0.0142003 & 0.00574549 \\
 C_9 & -0.0568647 & -0.124537 & 0.36619 & 0.172665 & 0.0110234 & 0.006496 \\
 C_{10} & 0.0538771 & 0.180566 & -0.168788 & -0.397328 & 0.0081847 & 0.00364168 \\
 C_{11} & -0.0307483 & -0.0719283 & 0.251402 & 0.146651 & 0.0061147 & 0.00323952 \\
 C_{12} & 0.0302112 & 0.107503 & -0.117035 & -0.286246 & 0.00503853 & 0.00235587 \\
 C_{13} & -0.0170692 & -0.039884 & 0.166517 & 0.102934 & 0.00352374 & 0.00168728 \\
 C_{14} & 0.0174768 & 0.0647301 & -0.0759667 & -0.194213 & 0.00329114 & 0.00158749 \\
 C_{15} & -0.00960039 & -0.0210649 & 0.10914 & 0.0648543 & 0.0020084 & 0.000850251 \\
 C_{16} & 0.0102464 & 0.0391984 & -0.0474513 & -0.126728 & 0.00230035 & 0.00114293 \\
 C_{17} & -0.00529678 & -0.00998488 & 0.0709918 & 0.0361078 & 0.00102572 & 0.000338474 \\
 C_{18} & 0.00590122 & 0.0231947 & -0.0281582 & -0.0782416 & 0.00175917 & 0.000906316 \\
 C_{19} & -0.00265197 & -0.00304648 & 0.0448996 & 0.0147282 & 0.000307947 & -0.0000173967 \\
 C_{20} & 0.00305022 & 0.0121834 & -0.0143011 & -0.0413795 & 0.00147647 & 0.000784372 \\
 C_{21} & -0.000700966 & 0.00224554 & 0.0247396 & -0.00338177 & -0.000177699 & -0.00022083\\
 \hline
  \hline
\end{array}\]
\end{table*}

 \begin{table}[htbp]
 \small
 \renewcommand{\arraystretch}{1.2}
\centering
\caption{The decay information of  $K_0^*$\text{(1430)}. The unit of the width is MeV. \label{1P}}
\begin{tabular}{ccccc}
\hline
 \hline
 \text{channel} & \text{width} & \text{BR}&\text{Exp. \cite{ParticleDataGroup:2024cfk}}\\
  \midrule[0.7pt]  
 $ K \pi$ & 252 & 0.955&$ 0.93\pm0.1$\\
 $\eta K$ & 11.1 & 0.042&$ 0.086^{+0.037}_{-0.034}$\\
   \text{Total width} & 264 & &$270\pm80$\\
  \hline
 \hline
\end{tabular}
\end{table}
Table \ref{nP} provides the two-body strong decay information for the excited kaonic states: $K_0^*(4P)$, $\kappa(2600)(5P)$, and $K_0^*(6P)$. These decays are crucial {to} understanding the resonance structures.

\begin{table*}[htbp]
 \small
 \renewcommand{\arraystretch}{0.86}
\centering
\caption{The {two-body} strong decay information of $K_0^*(4P)$ state, $\kappa(2600)(5P)$ state, and $K_0^*(6P)$   state. The unit of the width is MeV.  {Channels} with {branching} ratios less than 0.2\% {are omitted}. \label{nP}}
\vspace{-10pt}
\[\begin{array}{ccc}
\hline
\hline
\begin{array}{ccc}
&K_0^*(4P)&\\
  \midrule[0.7pt]  
 \text{channel} & \text{width} & \text{BR} \\
 K_1\pi  & 49.4 & 0.104 \\
{\pi K(4S)} & 38.7 & 0.0815 \\
K \pi & 29.1 & 0.0611 \\
 K^*\text{(1410)$\rho $} & 27.3 & 0.0575 \\
 \pi K_1\text{(2P)} & 27.1 & 0.057 \\
 \text{K$\pi $(1300)} & 20.8 & 0.0438 \\
 \text{$\pi $K(3S)} & 19.3 & 0.0406 \\
 Ka_1\text{(1640)} & 19.1 & 0.0402 \\
 \text{$\pi $K(1460)} & 17.4 & 0.0367 \\
 K^*\rho  & 16 & 0.0338 \\
 Ka_1 & 14.5 & 0.0304 \\
 \pi K_2\text{(2D)} & 13.6 & 0.0286 \\
 \pi K_1\text{(3P)} & 13.3 & 0.0281 \\
 K\pi _2\text{(1880)} & 13.2 & 0.0277 \\
 \pi K_1^\prime\text{(3P)} & 11.4 & 0.0239 \\
 Kb_1\text{(2P)} & 10.3 & 0.0217 \\
 h_1\text{(1595)K} & 10.3 & 0.0216 \\
 K^*\text{$\rho $(1450)} & 9.55 & 0.0201 \\
 K^*a_1 & 9.18 & 0.0193 \\
 {K\pi (1800)} & 8.77 & 0.0185 \\
 Kb_1 & 8.37 & 0.0176 \\
 \omega K^*\text{(1410)} & 8.24 & 0.0173 \\
 \pi K_3\text{(1F)} & 6.02 & 0.0127 \\
 K_1^\prime\rho  & 5.41 & 0.0114 \\
 \omega K^* & 5.05 & 0.0106 \\
 K^*f_1 & 4.91 & 0.0103 \\
{h_1K }&{4.32} & {0.0091} \\
 {\pi K_2'\text{(2D)} }&{ 4.11 }&{ 0.00864 }\\
{ Kf_1\text{(2P)} }&{ 3.35 }&{ 0.00704 }\\
 {\text{$\eta $K(1460)} }&{ 3.33 }&{ 0.00701 }\\
 {K\rho _2\text{(1D)} }&{ 3.26 }&{ 0.00686 }\\
 {\pi K_1^\prime\text{(2P)} }&{ 3.04 }&{ 0.00639 }\\
 {\text{$\omega $(1420)}K^* }&{ 3.01 }&{ 0.00634 }\\
 {{\eta (1475)K} }&{ 2.95 }&{ 0.00621 }\\
 {f_1{(1420)K} }&{ 2.79 }&{ 0.00587 }\\
 {K^*f_0\text{(1500)} }&{ 2.32 }&{ 0.00488 }\\
 {\eta K_1 }&{ 2.26 }&{ 0.00476 }\\
 {\rho K_0^*\text{(1430)} }&{ 2.16 }&{ 0.00454 }\\
 {{\eta K} }&{ 1.94 }&{ 0.00408 }\\
 {\omega K_1^\prime }&{ 1.82 }&{ 0.00383 }\\
 {f_1K }&{ 1.77 }&{ 0.00373 }\\
{ K^*a_2 }&{ 1.74 }&{ 0.00367 }\\
 {\rho K_2^* }&{ 1.3 }&{ 0.00274 }\\
 {{\eta^\prime K} }&{ 1.21 }&{ 0.00256 }\\
 {K\omega _2\text{(1D)} }&{ 1.09 }&{ 0.00229 }\\
{ K^*\phi  }&{ 1.06 }&{ 0.00224 }\\
{ {\eta^\prime K(1460)} }&{ 1.02 }&{ 0.00215 }\\
&\cdots&\\
 &&\\
  &&\\
   &&\\
    &&\\
    &&\\
  &&\\
   &&\\
    &&\\&&\\
  &&\\
   &&\\
    &&\\
        &&\\&&\\
  &&\\
   &&\\
    &&\\
        &&\\&&\\
  &&\\
   &&\\
    &&\\
    &&\\
  &&\\
   &&\\
    &&\\
    \hline
 \text{Total width} & 475 & 1 \\
\end{array}
 & 
\begin{array}{ccc}
&\kappa(2600)(5P)&\\
  \midrule[0.7pt]  
 \text{channel} & \text{width} & \text{BR} \\
 K_1\pi  & 32.4 & 0.0751 \\
 K^*\text{(1410)$\rho $} & 27.6 & 0.0639 \\
 \pi K_1\text{(2P)} & 20.1 & 0.0466 \\
 Kb_1\text{(1960)} & 18.5 & 0.0428 \\
 {K \pi } & 17.1 & 0.0395 \\
 \pi K_1\text{(3P)} & 14.7 & 0.0341 \\
 \text{$\pi $K(5S)} & 14.5 & 0.0335 \\
 \pi K_2\text{(2D)} & 14.1 & 0.0327 \\
 Ka_1\text{(1640)} & 13.7 & 0.0318 \\
 {K\pi (1300)} & 13.1 & 0.0304 \\
 {\pi K(4S)} & 13.1 & 0.0302 \\
 \text{$\pi $K(1460)} & 11.5 & 0.0265 \\
 \pi K_1^\prime\text{(4P)} & 11 & 0.0254 \\
 K^*a_1\text{(1640)} & 10.9 & 0.0254 \\
 {\pi K(3S)} & 9.85 & 0.0228 \\
 K\pi _2\text{(1880)} & 9.84 & 0.0228 \\
 \omega K^*\text{(1410)} & 8.66 & 0.0201 \\
 {K\pi (1800)} & 8.46 & 0.0196 \\
 Ka_1 & 7.74 & 0.0179 \\
 \pi K_3\text{(1F)} & 7.24 & 0.0168 \\
 Kb_1\text{(2P)} & 6.89 & 0.016 \\
 Ka_3\text{(2030)} & 6.77 & 0.0157 \\
{ h_1 (1595)K} & 6.31 & 0.0146 \\
 K^*\rho  & 6.1 & 0.0141 \\
 K_1b_1 & 6.08 & 0.0141 \\
 K^*a_1 & 5.25 & 0.0122 \\
 \pi K_3^\prime\text{(2F)} & 4.89 & 0.0113 \\
 K^*\text{$\rho $(1450)} & 4.44 & 0.0103 \\
 \pi K_1^\prime\text{(3P)} & 4.35 & 0.0101 \\ 
 {h_1{(1965)K} }&{ 4.33 }&{ 0.01 }\\ {
 \pi K_2^\prime\text{(3D)} }&{ 4.29 }&{ 0.00993 }\\ {
 \rho K_1^\prime\text{(2P)} }&{ 3.91 }&{ 0.00905 }\\ {
 K^*\text{(1680)$\rho $} }&{ 3.65 }&{ 0.00844 }\\ {
 K_1h_1 }&{ 3.56 }&{ 0.00824 }\\ {
 Kb_1 }&{ 3.49 }&{ 0.00808 }\\ {
 K_1^\prime\rho  }&{ 2.92 }&{ 0.00676 }\\ {
 K^*f_1 }&{ 2.91 }&{ 0.00674 }\\ {
 \eta K(1460)}&{ 2.82 }&{ 0.00652 }\\ {
 Kf_1\text{(2P)} }&{ 2.76 }&{ 0.00638 }\\ {
 K_1a_1 }&{ 2.52 }&{ 0.00584 }\\ {
 K\rho _2\text{(1D)} }&{ 2.1 }&{ 0.00487 }\\ {
 h_1K }&{ 2.08 }&{ 0.00482 }\\ {
 K_1\text{$\pi $(1300)} }&{ 2.06 }&{ 0.00476 }\\ {
 \omega K^* }&{ 1.91 }&{ 0.00442 }\\ {
 {\eta K(3S)} }&{ 1.88 }&{ 0.00435 }\\ {
 \pi K_2^\prime\text{(2D)} }&{ 1.73 }&{ 0.00401 }\\ {
 Kf_3\text{(2050)} }&{ 1.73 }&{ 0.00401 }\\ {
 K^*a_2 }&{ 1.6 }&{ 0.0037 }\\ {
 Ka_1\text{(3P)} }&{ 1.55 }&{ 0.00358 }\\ {
 K^*a_0 }&{ 1.54 }&{ 0.00357 }\\ {
 f_1{(1420)K} }&{ 1.54 }&{ 0.00356 }\\ {
 K^*\text{(1410)$\phi $} }&{ 1.44 }&{ 0.00333 }\\ {
 \eta K_1 }&{ 1.42 }&{ 0.00329 }\\ {
 {\eta (1295)}K_1 }&{ 1.42 }&{ 0.00328 }\\ {
 {\eta K} }&{ 1.38 }&{ 0.00319 }\\ {
 {\eta (1475)K} }&{ 1.37 }&{ 0.00317 }\\ {
 \rho K_2^* }&{ 1.33 }&{ 0.00307 }\\ {
 \omega K_1^\prime\text{(2P)} }&{ 1.26 }&{ 0.00291 }\\ {
 K\rho _2\text{(2D)} }&{ 1.17 }&{ 0.00271 }\\ {
 \pi K_1^\prime\text{(2P)} }&{ 1.14 }&{ 0.00263 }\\ {
 \rho K_0^*\text{(1430)} }&{ 1.13 }&{ 0.00261 }\\ {
 \omega K^*\text{(1680)} }&{ 1.11 }&{ 0.00257 }\\ {
 K_1\rho  }&{ 1.02 }&{ 0.00235 }\\ {
 f_1K }&{ 1.01 }&{ 0.00234 }\\ {
 \omega K_1^\prime }&{ 0.996 }&{ 0.00231 }\\ {
 \text{$\omega $(1420)}K^* }&{ 0.982 }&{ 0.00228 }\\ {
 K^*\rho _2\text{(1D)} }&{ 0.956 }&{ 0.00221 }\\ {
 \rho K_2\text{(1770)} }&{ 0.93 }&{ 0.00215 }\\ 
 &\cdots&\\
 &&\\
  &&\\
   &&\\
    &&\\
    &&\\
    \hline
 \text{Total width} & 432 & \text{1} \\
\end{array}
 & 
\begin{array}{ccc}
&K_0^*(6P)&\\
  \midrule[0.7pt]  
 \text{channel} & \text{width} & \text{BR} \\
 K^*\text{(1410)$\rho $} & 21.6 & 0.0568 \\
 K_1\pi  & 20.8 & 0.0548 \\
 \pi K_1\text{(2P)} & 14.6 & 0.0384 \\
 Kb_1\text{(1960)} & 13.7 & 0.036 \\
 \pi K_1\text{(3P)} & 11.8 & 0.0312 \\
 \pi K_2\text{(2D)} & 11.6 & 0.0305 \\
 {K\pi} & 10.8 & 0.0283 \\
 {\pi K(5S)} & 9.69 & 0.0255 \\
 Ka_3\text{(2030)} & 9.68 & 0.0255 \\
 \pi K_1^\prime\text{(5P)} & 8.95 & 0.0236 \\
 {K\pi (1300)} & 8.87 & 0.0233 \\
 \pi K_2\text{(4D)} & 8.63 & 0.0227 \\
 {\pi K(1460)} & 8.16 & 0.0215 \\
 Ka_1\text{(1640)} & 8.13 & 0.0214 \\
 \pi K_1\text{(4P)} & 7.63 & 0.0201 \\
 {\pi K(4S)} & 7.39 & 0.0194 \\
 {K\pi(1800)} & 7.31 & 0.0193 \\
 K^*a_1\text{(1640)} & 7.29 & 0.0192 \\
 \pi K_3^\prime\text{(2F)} & 7.25 & 0.0191 \\
 \omega K^*\text{(1410)} & 6.87 & 0.0181 \\
 \pi K_3\text{(1F)} & 6.4 & 0.0169 \\
 K\pi _2\text{(1880)} & 6.2 & 0.0163 \\
 {\pi K(3S)} & 5.89 & 0.0155 \\
 K_1b_1 & 5.64 & 0.0148 \\
 \pi K_1^\prime\text{(4P)} & 5.3 & 0.0139 \\
 K^*\text{(1410)}a_1 & 4.76 & 0.0125 \\
 Kb_1\text{(2P)} & 4.67 & 0.0123 \\
 h_1{(1595)K} & 4.32 & 0.0114 \\
 Ka_1 & 4.18 & 0.011 \\
{\pi K_1^\prime\text{(3P)} }&{ 2.19 }&{ 0.00577 }\\ {
 K_1h_1 }&{ 2.17 }&{ 0.00572 }\\ {
 {\eta K(3S)} }&{ 2.16 }&{ 0.00569 }\\ {
 K^*f_1\text{(2P)} }&{ 2.09 }&{ 0.00549 }\\ {
 K^*\text{$\rho $(1450)} }&{ 2.08 }&{ 0.00548 }\\ {
 {\eta K(1460)} }&{ 2 }&{ 0.00526 }\\ {
 K^*\rho _2\text{(1D)} }&{ 1.89 }&{ 0.00498 }\\ {
 \rho K^*\text{(3S)} }&{ 1.87 }&{ 0.00493 }\\ {
 h_1{(2215)K} }&{ 1.76 }&{ 0.00464 }\\ {
 Kf_1\text{(2P)} }&{ 1.68 }&{ 0.00441 }\\ {
 Kb_1 }&{ 1.66 }&{ 0.00436 }\\ {
 K^*f_1 }&{ 1.65 }&{ 0.00434 }\\ {
 K^*a_0\text{(1740)} }&{ 1.62 }&{ 0.00427 }\\ {
 Kb_1\text{(2240)} }&{ 1.62 }&{ 0.00427 }\\ {
 b_1{K(1460)} }&{ 1.61 }&{ 0.00423 }\\ {
 K_1^\prime\rho  }&{ 1.58 }&{ 0.00416 }\\ {
 {\pi (1300)K(1460)} }&{ 1.52 }&{ 0.004 }\\ {
 K^*a_2 }&{ 1.46 }&{ 0.00384 }\\ {
 K_1^\prime a_1 }&{ 1.44 }&{ 0.00379 }\\ {
 \rho K_1^\prime\text{(2P)} }&{ 1.42 }&{ 0.00375 }\\ {
 K\rho _2\text{(2D)} }&{ 1.42 }&{ 0.00374 }\\ {
 \pi K_4\text{(1G)} }&{ 1.27 }&{ 0.00334 }\\ {
 \rho K_1\text{(2P)} }&{ 1.22 }&{ 0.00322 }\\ {
 K\pi _2\text{(3D)} }&{ 1.2 }&{ 0.00317 }\\ {
 \rho K_2\text{(1770)} }&{ 1.15 }&{ 0.00303 }\\ {
 \omega K^*\text{(1680)} }&{ 1.12 }&{ 0.00296 }\\ {
 \eta K_1 }&{ 1.08 }&{ 0.00285 }\\ {
 K^*a_0 }&{ 1.08 }&{ 0.00285 }\\ { h_1K }&{ 1.08 }&{ 0.00285 }\\ {
 {K\eta (4S)} }&{ 1.04 }&{ 0.00274 }\\ {
 K^*\text{(1410)}f_1 }&{ 1.02 }&{ 0.0027 }\\ {
 K\rho _2\text{(1D)} }&{ 1.02 }&{ 0.00269 }\\ {
 Kf_1\text{(3P)} }&{ 0.939 }&{ 0.00247 }\\ {
 K^*f_2\text{(2P)} }&{ 0.938 }&{ 0.00247 }\\ {
 f_1{(1420)K} }&{ 0.934 }&{ 0.00246 }\\ {
 K_1\rho  }&{ 0.927 }&{ 0.00244 }\\ {
 K^*\text{$\rho $(1700)} }&{ 0.894 }&{ 0.00235 }\\ {
 \rho K_2^* }&{ 0.887 }&{ 0.00233 }\\ {
 \omega K^* }&{ 0.885 }&{ 0.00233 }\\ {
 \pi K_2^\prime\text{(2D)} }&{ 0.87 }&{ 0.00229 }\\ {
 {\eta K} }&{ 0.832 }&{ 0.00219 }\\ {
 {\eta (1475)K} }&{ 0.799 }&{ 0.0021 }\\ {
 K^*\rho _3\text{(1690)} }&{ 0.797 }&{ 0.0021 }\\ {
 \rho K_2^*\text{(1980)} }&{ 0.788 }&{ 0.00207 }\\ 
 &\cdots&\\
 \hline
 \text{Total width} & 380 & \text{1} \\
\end{array}
 \\
 \hline\hline
\end{array}\]
\end{table*}

Reference \cite{Li:2022ybj} also calculated the two-body strong decays of $K_0^*$ meson family using the QPC model. In their approach, the wave functions were derived from the GI potential model, with the parameter $\gamma$ fixed at 0.52. In contrast, the value of $\gamma$ used in our work differs by a factor of $\sqrt{96\pi}$~\cite{Song:2014mha}, which leads to different results for the two-body strong decays of $K_0^*(4P)$. And the different $\gamma$ value would cause our overall results to be 1.3 times larger than those in Ref.~\cite{Li:2022ybj}.

The primary decay channel of $K_0^*(4P)$ is $K_1\pi$, with a width of about 49 MeV, contributing 10.4\% to the total width, which is much larger than the one obtained by Ref. \cite{Li:2022ybj}. 
Other significant decay modes include $\pi K(4S)$ and $K\pi$, which contribute 8.15\%
and 6.11\%, respectively. 
$K\pi$ will be an ideal final channel to search {for} the $K_0^*(4P)$ state.
$K^*(1410)\rho$ and $\pi K_1(2P)$ also have larger {contributions} to the total width about 5.7\%. $K^*(1410)\rho$ has {a} width of about 33 MeV in Ref. \cite{Li:2022ybj}. {Other} decay channels with branching ratios less than 5\% can be found in Table \ref{nP}. 
In the QPC model, the processes $K_0^*(4P)\to {K^*(1410)\omega}$ and $K_0^*(4P)\to {K^*(1410)\rho}$ share the same initial state $K_0^*(4P)$ and the $K^*(1410)$ {final-state component}. 
The wave functions of $\omega$ and $\rho$ are the same except {for} the flavor wave functions. Combining the relationship $m_{\omega}\simeq m_{\rho}$, their decay outcomes depend on the overlap of the flavor matrix elements.
Finally, {we can roughly obtain $\frac{\Gamma_{K^*(1410)\rho}}{\Gamma_{K^*(1410)\omega}}\simeq 3$.
In our calculation, $\frac{\Gamma_{K^*(1410)\rho}}{\Gamma_{K^*(1410)\omega}}\simeq 3.31$. Another 
 relation, $\frac{\Gamma_{K^*\rho}}{\Gamma_{K^*\omega}}\simeq 3.17$,} is also  {consistent} with 3.
 {Other} decay channels can be found in Table \ref{nP}. 
Due to differences in both the model parameter $\gamma$ and the choice of wave functions, our calculated total width of $475\,\text{MeV}$ for the $K_0^*(4P)$ state is about 2 times larger than that reported in Ref. \cite{Li:2022ybj} when they take $M_{K_0^*(4P)}=2450$ MeV. In addition to the different choices of $\gamma$ and wave {functions}, we have also considered the higher excited final states which have not yet been found in experiment (such as $\pi K(3S)$ and $\pi K(4S)$.

For $\kappa(2600)(5P)$,  the largest decay channel is also $K_1\pi$, accounting for 7.5\% of the total width. $K^*(1410)\rho$ is comparably significant (6.39\%). This state also exhibits considerable decay fractions into higher  {excitations}, such as $\pi K_1(2P)$ and $K b_1(1960)$, {each contributing}  larger than 4\%.  {The} $K\pi$ channel constitutes a major decay mode of {the} $\kappa(2600)(5P)$ state, consistent with its experimental observation in the same channel. $\pi K_1(3P)$,  $\pi K(5S)$, $\pi K_2(2D)$, $K a_1(1640)$, and $\pi K(4S)$,  all {have a significant contribution} to the total width of $\kappa(2600)(5P)$, {with {branching} ratios being all about} 3\%. {Other} decay channels of $\kappa(2600)(5P)$ {with} branching ratios less than 3\% can be found in Table \ref{nP}. The important thing is the total decay width of $\kappa(2600)(5P)$ 
 {overlaps} with the experimental value $480\pm47\pm72$ MeV. 
This result supports our classification of  $\kappa(2600)$  as  the  $K_0^*(5P)$ {resonance}.

$K_0^*(6P)$ configuration demonstrates remarkable parallels to $K_0^*(5P)$, with the primary eight decay channels showing negligible divergence. 
The most important decay modes include $K^*(1410)\rho$ (5.7\%) and $K_1\pi$ (5.5\%).
$\pi K_1(2P)$,  $Kb_1(1960)$, $\pi K_1(3P)$, and $\pi K_2(2D)$ are its larger decay channels with {branching} ratios of $3.1\%$ to $3.8\%$.  As in $K_0^*(5P)$,  the $K\pi$ channel constitutes a major decay mode of this  $K_0^*(6P)$ state.  Experiments can search for this state through {the} $K\pi$ final channel.
This state exhibits a more evenly distributed decay pattern, with multiple channels contributing between  {2\% and 4\%}. 
Other decay channels of $K_0^*(6P)$  with more details are listed in Table \ref{nP}. 
{We find that} the total width of $K_0^*(6P)$ is 380 MeV.


\section{CONCLUSION}\label{sec4}
In this work, we have systematically studied the mass spectra and the OZI-allowed two-body strong decay behaviors of the newly observed $\kappa(2600)$, along with the higher excited kaonic states, $K_0^*(4P)$ and $K_0^*(6P)$. 
\par

The newly observed state  $\kappa(2600)$ is a {good} candidate for the $K_0^*(5P)$ state. The findings are supported not only by the Regge trajectory and the MGI model results, but also through an analysis of the QPC model. The total and branching decay widths allow for the identification of $\kappa(2600)$ as the $K_0^*(5P)$ state. Its two most important decay channels are $K_1\pi$ and $K^*(1410)\rho$.

The masses of $K_0^*(4P)$  and $K_0^*(6P)$ states {are} predicted to be about 2450 MeV and 2830 MeV, respectively. The total widths for the $K_0^*(4P)$ state and $K_0^*(6P)$ state are about 475 MeV and 380 MeV, respectively. The  $K\pi$ channel is an ideal final channel for searching these two $K_0^*$ states.  $K_1\pi$ and $\pi K(4S)$ are two of the most important decay channels of the $K_0^*(4P)$ state, with {branching} ratios of 10.4\% and 8.2\% respectively. The  {branching} ratio of $K\pi$ channel is about 6\% for the $K_0^*(4P)$ state.  The $K^*(1410)\rho$  and $K_1\pi$ channels are two main final states of the  $K_0^*(6P)$ state, and they have {branching} ratios larger than 5\%.

We look forward to future experimental studies, which will play a crucial role in searching for the higher excited states of kaons.

\begin{acknowledgments}
This work is supported by the National Natural Science Foundation of China under Grants  No.~11965016 and No.~12247101, and by the Natural Science Foundation of Qinghai Province under Grant No.~2022-ZJ-939Q, the Fundamental Research Funds for the Central Universities (Grant No.~lzujbky-2024-jdzx06).
\end{acknowledgments}

\section*{Data availability}
The data that support the findings of this article are openly available~\cite{pang2025dataset}.

\newpage

\bibliographystyle{apsrev4-1}
%
\end{document}